\documentclass[runningheads]{llncs}
\usepackage{graphicx}
\usepackage{comment}
\usepackage{amsmath,amssymb} 
\usepackage{color}
\usepackage{float}
\usepackage{multirow}
\usepackage{subfig}
\usepackage{xcolor}
\definecolor{mypink1}{rgb}{0.858, 0.188, 0.478}

\begin{document}
\pagestyle{headings}
\mainmatter
\def\ECCVSubNumber{073}  

\title{AWNet: Attentive Wavelet Network for Image ISP} 

\titlerunning{AWNet: Attentive Wavelet Network for Image ISP}
%
\author{Linhui Dai\inst{\thanks{Authors contributed equally}}, 
Xiaohong Liu\inst{\footnotemark[1]} \and
Chengqi Li \and
Jun Chen}
\authorrunning{Dai et al.}
%
\institute{McMaster University, Ontario, Canada \\
\email{\{dail5, liux173, lic222, chenjun\}@mcmaster.ca}}

\maketitle
\begin{abstract}
	As the revolutionary improvement being made on the performance of smartphones over the last decade, mobile photography becomes one of the most common practices among the majority of smartphone users. However, due to the limited size of camera sensors on phone, the photographed image is still visually distinct to the one taken by the digital single-lens reflex (DSLR) camera. To narrow this performance gap, one is to redesign the camera image signal processor (ISP) to improve the image quality. Owing to the rapid rise of deep learning, recent works resort to the deep convolutional neural network (CNN) to develop a sophisticated data-driven ISP that directly maps the phone-captured image to the DSLR-captured one. In this paper, we introduce a novel network that utilizes the attention mechanism and wavelet transform, dubbed AWNet, to tackle this learnable image ISP problem. By adding the wavelet transform, our proposed method enables us to restore favorable image details from RAW information and achieve a larger receptive field while remaining high efficiency in terms of computational cost. The global context block is adopted in our method to learn the non-local color mapping for the generation of appealing RGB images. More importantly, this block alleviates the influence of image misalignment occurred on the provided dataset. Experimental results indicate the advances of our design in both qualitative and quantitative measurements. The source code is available at \textcolor{mypink1}{https://github.com/Charlie0215/AWNet-Attentive-Wavelet-Network-for-Image-ISP}.
\end{abstract}

\section{Introduction}
    Traditional image ISP is a critical processing unit that maps RAW images from the camera sensor to RGB images in order to accommodate the human visual system (HVS). For this purpose, a series of sub-processing units are leveraged in order to tackle the different artifacts from photo-capturing devices, including, among others, the color shifts, signal noises, and moire effects. However, tuning each sub-processing unit requires legions of efforts from imagery experts.
    
    Nowadays, mobile devices have been equipped with high-resolution cameras to serve the incremental need for mobile photography. However, due to the compact space, the hardware is limited with respect to the quality of the optics and the pixel numbers. Moreover, the time of exposure is relatively short due to the instability of hand-holding. Therefore, a mobile specific ISP has to compensate for these limitations as well.
    
\begin{figure}[t]
	\centering
    \captionsetup[subfigure]{labelformat=empty}
    \begin{minipage}{0.45\textwidth}
        \vspace{2.1mm}
        \subfloat[RGB results of 1.png from our AWNet.]{\includegraphics[width=0.9\linewidth]{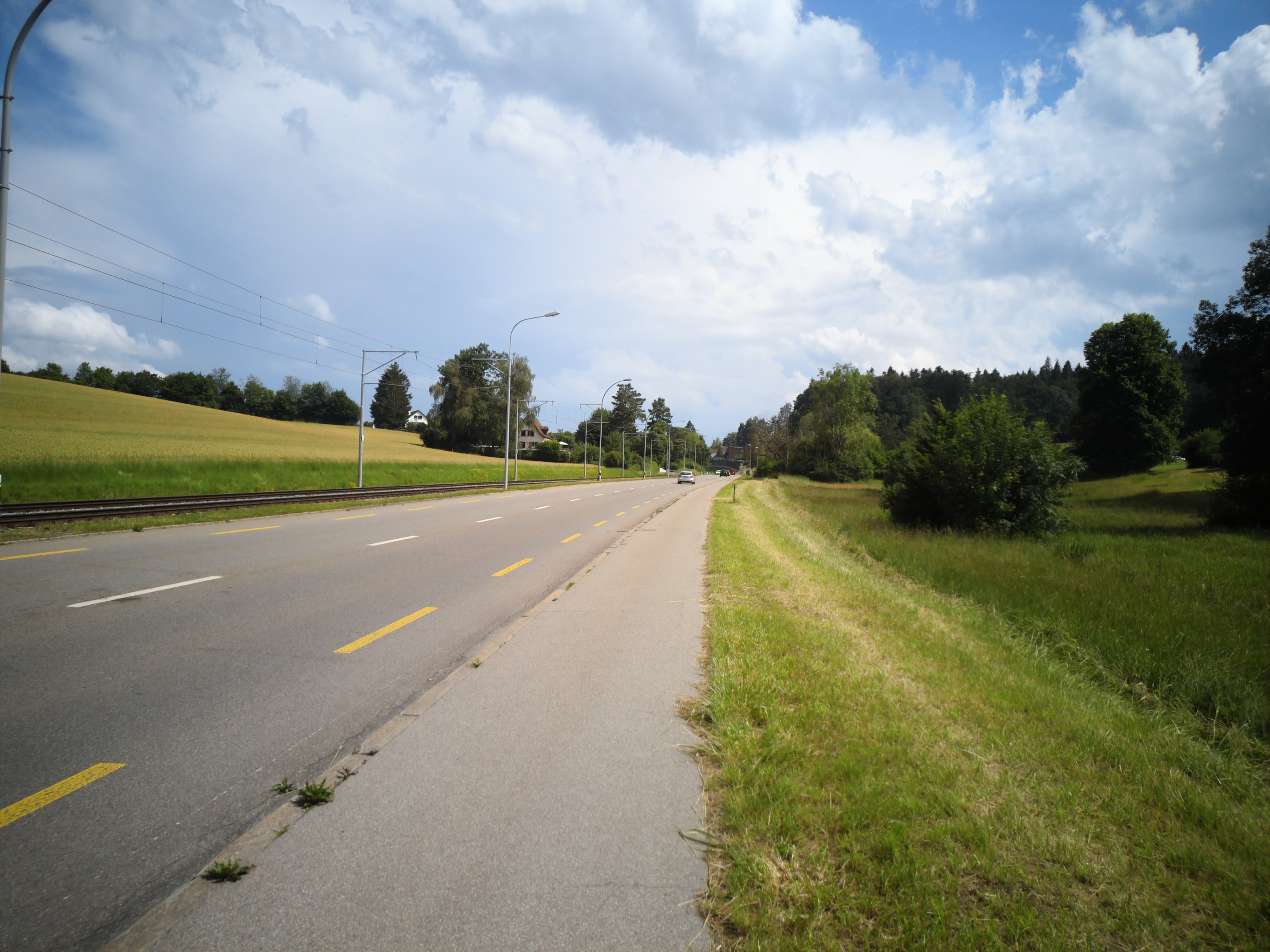}}
    \end{minipage}
    \hspace{2mm}
    \begin{minipage}{0.45\textwidth}
        \subfloat
        {
             
            \subfloat[R channel]{\includegraphics[width=0.3333\linewidth]{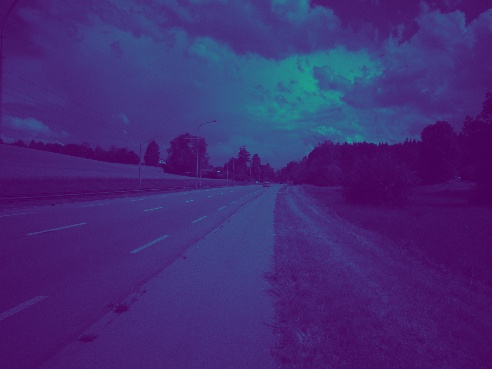}}\hspace{0.03\textwidth}
            \subfloat[G1 channel]{\includegraphics[width=0.3333\linewidth]{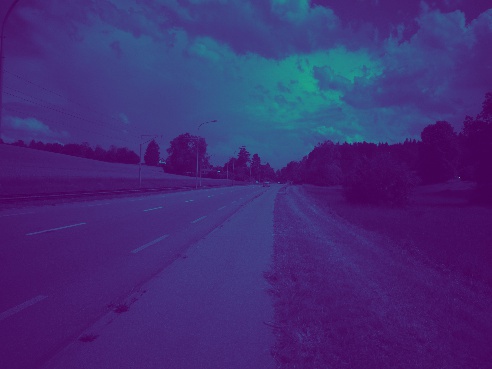}}
        }\\
        \subfloat
        {
            \subfloat[B channel]{\includegraphics[width=0.3333\linewidth]{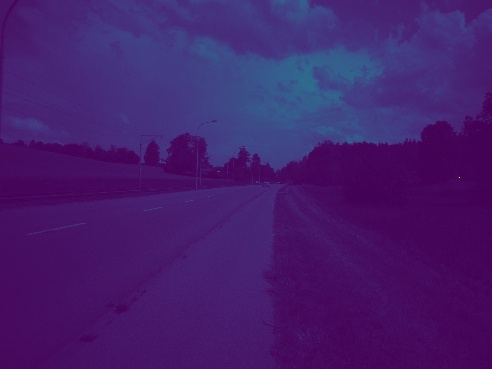}}\hspace{0.03\textwidth}
            \subfloat[G2 channel]{\includegraphics[width=0.3333\linewidth]{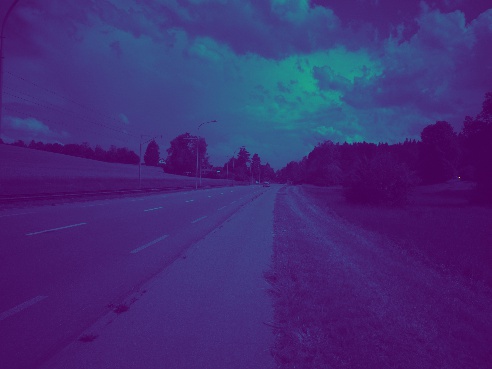}}
        }
        \vspace{2.1mm}
    \end{minipage}
    \caption{Visualization of each channel in the RAW image and the corresponding RGB image reconstructed by AWNet. Zoom-in for better views.}
    \label{fig:image1}
    \vspace{-2mm}
\end{figure}

    Recently, deep learning (DL) based methods have achieved considerable success on various image enhancement tasks, including image denoising \cite{abdelhamed2020ntire,zhang2017beyond}, image demosaicing \cite{gharbi2016deep}, and super-resolution \cite{kim2016accurate,ledig2017photo,lugmayr2020ntire,wang2019edvr}. Different from traditional image processing algorithms that commonly require prior knowledge of natural image statistics, data-driven methods can implicitly learn such information. Due to this fact, the DL-based method becomes a good fit for mapping problems \cite{chen2018learning,xu2019towards,zhu2017unpaired}.
    In here, learning image ISP can be regarded as an image-to-image translation problem, which can be well-addressed by the DL-based method. In ZRR dataset from \cite{ignatov2020replacing}, the RAW images can be decomposed into 4 channels, which are red (R), green (G1), blue (B) and green (G2) from the Bayer pattern, as shown in Fig.~\ref{fig:image1}. Remark that 2 of 4 channels record the radiance information from green sensors. Therefore, additional operations such as demosaicing and color correction are needed to tackle the RAW images as compared to RGB images. Moreover, due to the nature of the Bayer filter, the size of these 4 channels is down-sampled by the factor of two. In order to make the size of prediction and ground truth images consistent, an up-sampling operation is required. This can be regarded as a restoration problem, where the recovery of high-frequency information should be taken into consideration. In our observation, the misalignment between the DSLR and mobile photographed image pairs is severe even though the authors have adopted the SIFT \cite{lowe2004distinctive} and RANSAC \cite{vedaldi2010vlfeat} algorithms to mitigate this effect. It is worth mentioning that the minor misalignment between the input RAW image and ground-truth RGB image would cause a significant performance drop.
    
    To tackle the aforementioned problems, we introduce a novel trainable pipeline that utilizes the attention mechanism and wavelet transform. More specifically, the input of our proposed methods is a combination of a RAW image and its demosaiced counterpart as a complement, where the two-branch design is aimed at emphasizing the different training tasks, namely, noise removal and detail restoration on RAW model and the color mapping on the demosaiced model; the discrete wavelet transform (DWT) is adopted to restore fine context details from RAW images while reserving the informativeness in features during training; as for the color correction and tone mapping, the res-dense connection and attention mechanism are utilized to encourage the network putting effort on the focused areas.

    In summary, our main contributions are:
    \begin{itemize}
    \item[1)] Exploring the effectiveness of wavelet transform and non-local attention mechanism in image ISP pipeline.
    \item[2)] A two-branch design to take a raw image and its demosaiced counterpart that endows our proposed method the ability to translate the RAW image to the RGB image.
    \item[3)] A lightweight and fully convolutional encoder-decoder design that is time-efficient and flexible on different input sizes.
    \end{itemize}

\section{Related Works}
    In this section, we provide a brief review of the traditional image ISP methods, some representative RAW to RGB mapping algorithms, and the existing learnable imaging pipelines.
    \subsection{Traditional Image ISP Pipeline}
    Traditional ISP pipeline encompasses multiple image signal operations, including, among others, denoising, demosaicing, white balancing, color correction, gamma correction, and tone mapping. 
    Due to the nature of the image sensor, the existence of noise in RAW images is inevitable. Therefore, some operations are \cite{abdelhamed2020ntire,dabov2007image,zhang2017beyond} proposed to remove the noise and improve the signal-to-noise ratio.  The demosaicing operation interpolates the single-channel raw image with repeated mosaic
patterns into multi-channel color images \cite{gharbi2016deep}. White balancing corrects the color by shifting illuminations of RGB channels to make the image more perceptually accepted \cite{cheng2015beyond}. Color correction adjust the image value by a correction matrix \cite{kwok2013simultaneous,rizzi2003new}. Tone mapping shrinks the histogram of image values to enhance image details \cite{rana2019deep,yuan2012automatic}.
    Note that all sub-processing units in the traditional image ISP pipeline require human effort to manually adjust the final result.
    \subsection{RAW Data Usage in Low-level Image Restoration}
    The advantages of applying RAW data on low-level vision tasks have been explored by different works in the field of image restoration. For instance, \cite{chen2018learning} uses dark RAW image and bright color image pairs to restore dark images from images with long exposure. In this case, the radiance information that retained by raw data contributes to the restoration of image illumination.
    \cite{xu2019towards} takes advantage of rich radiance information from unprocessed camera data to restore high frequency details and improve their network performance on super-resolution tasks. Their experiment reveals that using raw data as a substitute for camera processed data is beneficial on single image super-resolution tasks.
    Lately, \cite{ignatov2020replacing,schwartz2018deepisp} adopt unprocessed image data to enhance mobile camera imaging. Since RAW data avoids the information loss introduced by quantization in ISP, it is favorable for a neural network to restore the delicate image details. Inspired by \cite{ignatov2020replacing}, our work makes use of the RAW data to train our network for a learnable ISP pipeline. Instead of only taking RAW images as the input, we adopt the combination of the input data formats from \cite{ignatov2020replacing} and \cite{schwartz2018deepisp} to encourage our network to learn different sub-tasks of image ISP, for example, noise removal, color mapping, and detail restoration. 
    
    \subsection{Deep Learning Based Image ISP Pipeline}
    Since CNN has achieved the promising performance on plenty of low-level vision tasks \cite{he2019mop,kim2016accurate,ledig2017photo,tao2018scale,wang2019edvr}, it is intuitive to leverage it for the learning of camera ISP. \cite{schwartz2018deepisp} collects RAW low-lit images from Samsung S7 phone, and uses a neural network to improve image brightness and remove noise on demosaiced RGB images from a simple ISP pipeline. \cite{ratnasingam2019deep} generates synthetic RAW images from JPEG ones and applies RAW-to-RGB mapping to restore the original RGB images. Moreover, some previous works in AIM 2019 RAW to RGB Mapping Challenge have achieved appealing results. For example, \cite{uhm2019w} considers using the stacked U-Nets to produce a pipeline in a coarse-to-fine manner. \cite{mei2019higher} adopts a multi-scale training strategy that recovers the image details while remaining the global perceptual acceptance. The most recent work \cite{ignatov2020replacing} tries to narrow the visual quality gap between the mobile and DSLR color images by directly translating mobile RAW images to DSLR color ones, where RAW images are captured by Huawei P20 phone and color ones are from Canon 5D Mark IV. Nonetheless, all previous learnable ISP methods only focus on the general mapping problem without mentioning other artifacts from the training dataset. For example, without additional operation, the misalignment between the DSLR and mobile image pairs can cause severe degradation on estimated outputs. In our work, we apply the global context block combined with the res-dense block that learns the global color mapping to tackle misaligned image features. The added blocks enable our network to outperform the current state-of-the-art method proposed by \cite{ignatov2020replacing}.
    \begin{figure}[h]
    \centering
        	\begin{minipage}[h]{0.9\linewidth}
        	\includegraphics[width=\linewidth]{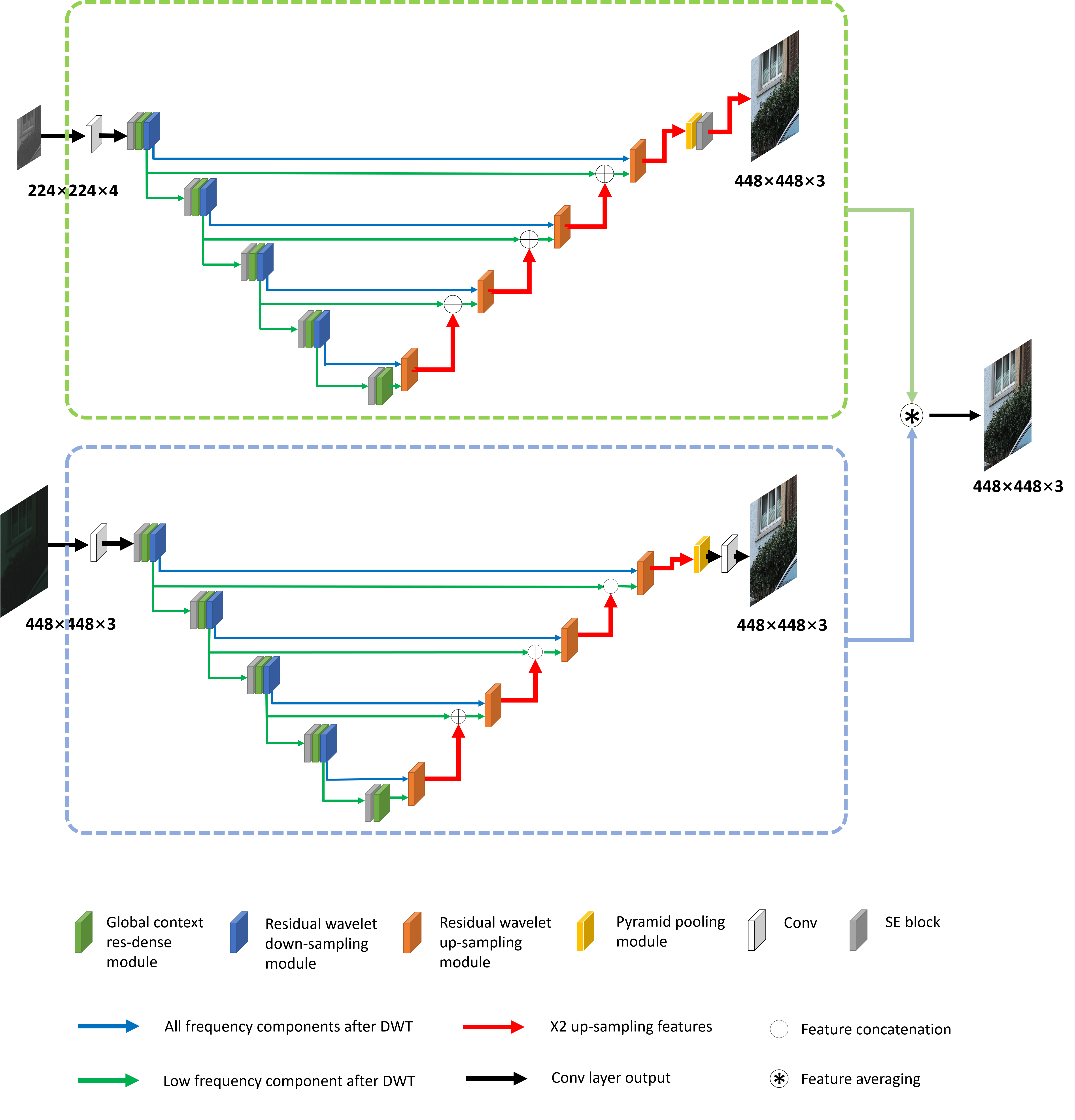}
            \end{minipage}
        \caption{The main architecture of the proposed AWNet. The top and bottom ones are the RAW and demosaiced models, respectively. We take the average of both outputs from these two models to obtain the final prediction.}
        \label{fig:AWNet}
    	\vspace{-2mm}
    \end{figure}     
    
    \begin{figure}[t]
    \centering
        	\begin{minipage}[h]{0.8\linewidth}
        	\includegraphics[width=\linewidth]{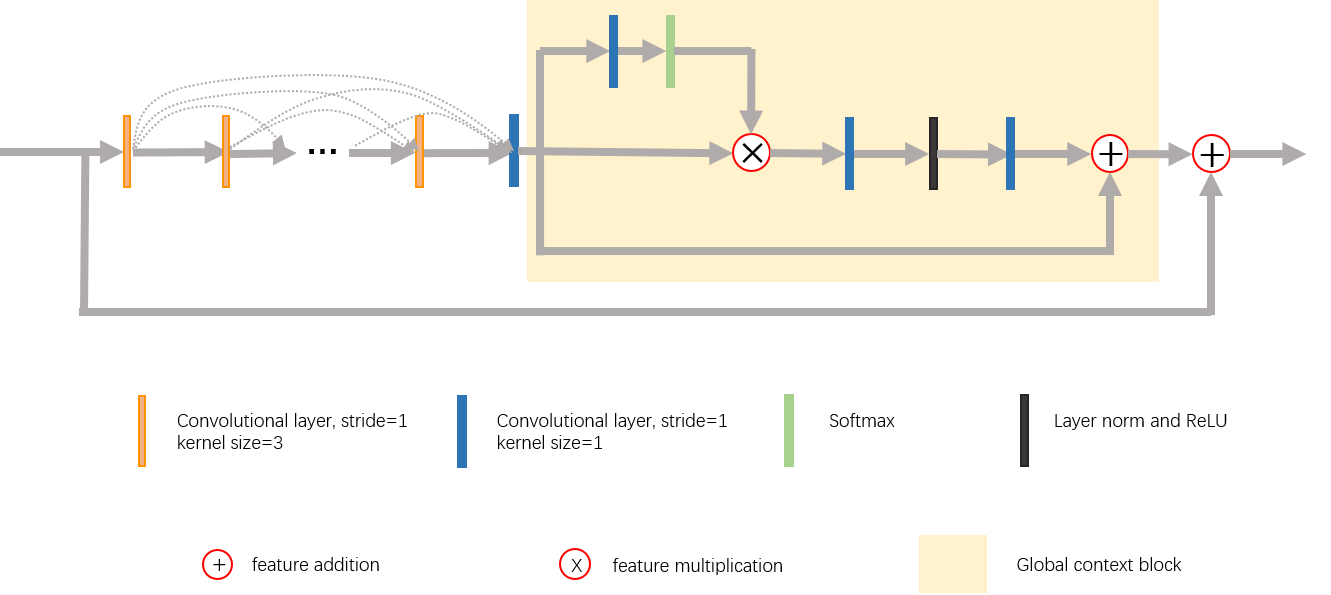}
            \end{minipage}
        \caption{Our global context res-dense module contains a residual dense block (RDB) and a global context block (GCB). We observe that the RDB can benefit the color restoration from RAW images and the GCB encourages the network putting effort on learning the global color mapping. See details in \ref{section:ablation}.}
        \label{fig:rdb}
    	\vspace{-2mm}
    \end{figure}   

    \begin{figure}[h]
    	\centering
    	\begin{minipage}[h]{0.4\linewidth}
    		\centering
    		\includegraphics[width=\linewidth]{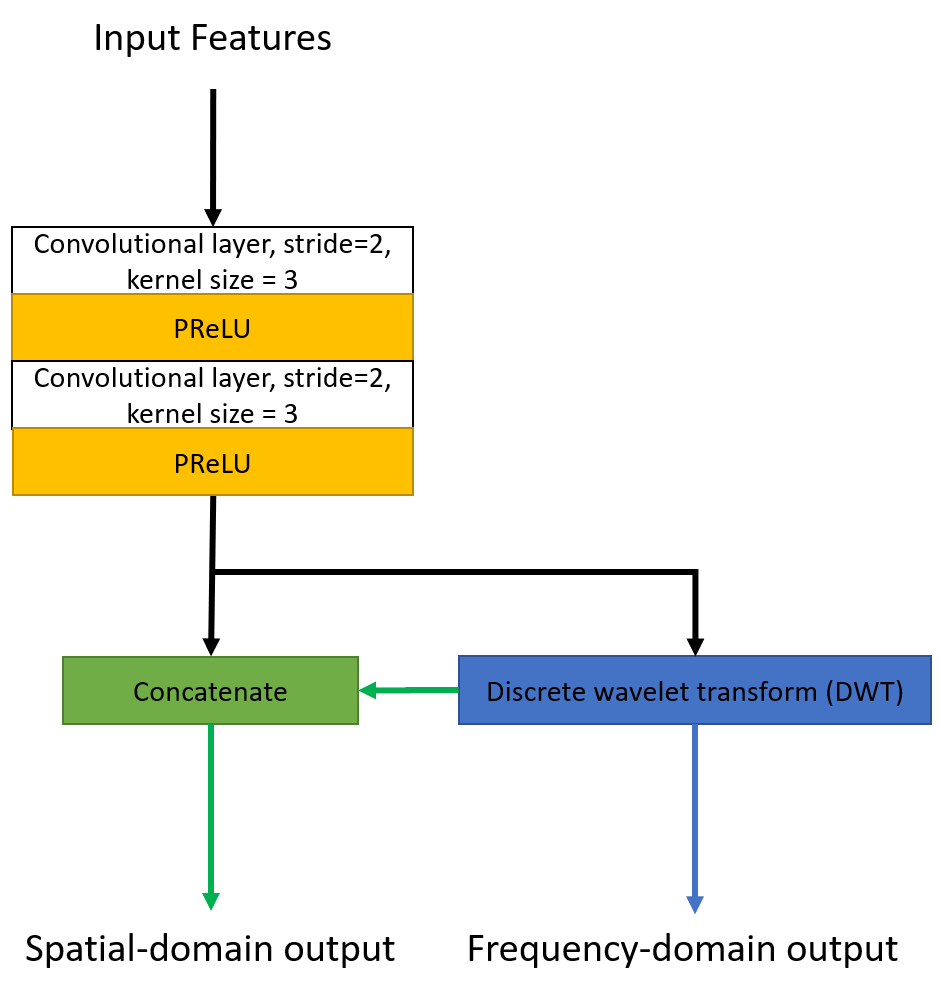}
    		\scriptsize{(a) Residual Wavelet Down-sampling Block}
    	\end{minipage}
    	\begin{minipage}[h]{0.4\linewidth}
    		\centering
    		\includegraphics[width=\linewidth]{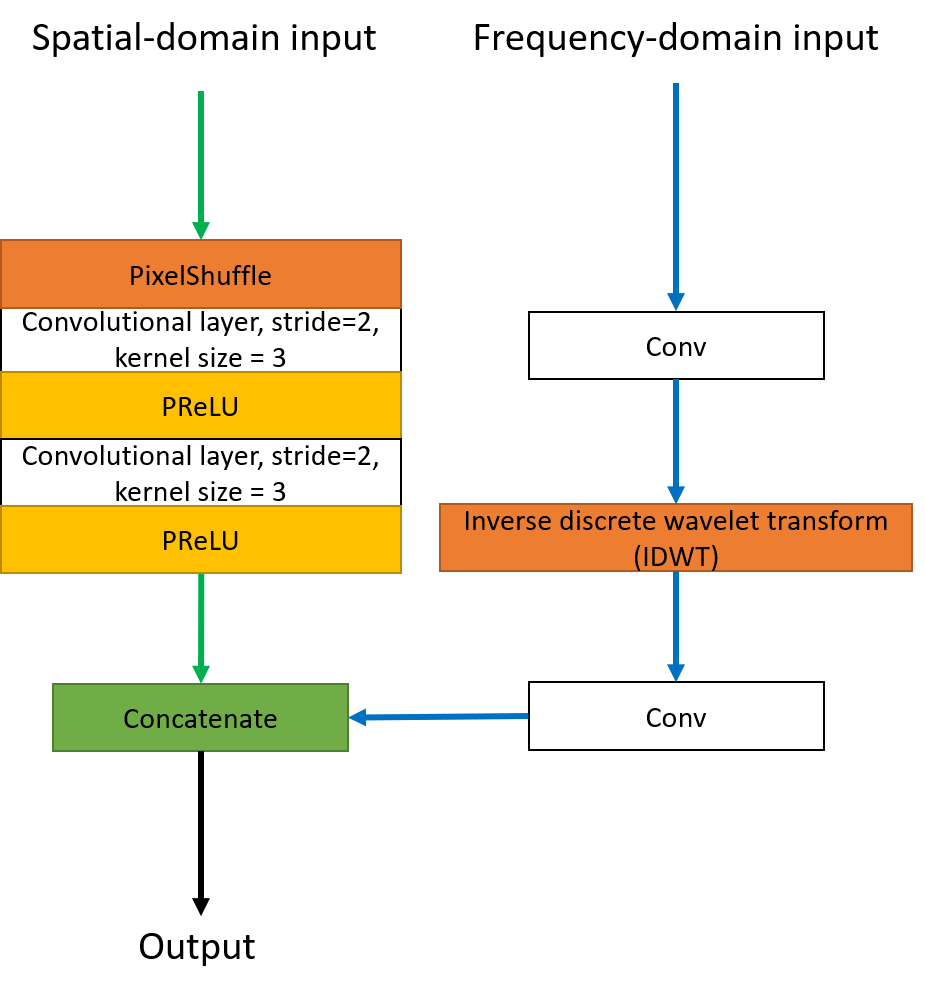}
    		\scriptsize{(b) Residual Wavelet Up-sampling Block}
    	\end{minipage}
    \caption{Illustration of our up-sampling and down-sampling modules in Fig.~\ref{fig:AWNet}. The residual design enables our model to operate in frequency-domain and spatial-domain that facilitates the learning of abundant features in up-sampling and down-sampling blocks.}
    	\label{fig:subblocks}
    \end{figure}
\section{Proposed Methods}
    We describe the proposed method and training strategy in this section. First, the overall network architecture (shown in Fig.~\ref{fig:AWNet}) and details of each network module are demonstrated, and then the sense of this design is illustrated. In the end, the loss functions adopted in training is introduced.
    
\subsection{Network Structure}
    The proposed AWNet employs a U-Net resembled structure and consolidates the architecture by three main modules, namely global context res-dense module, residual wavelet up-sampling module, and residual wavelet down-sampling module (see Fig.~\ref{fig:rdb} and Fig.~\ref{fig:subblocks}).
    
    The global context res-dense module consists of a residual dense block (RDB) and a global context block (GCB) \cite{cao2019gcnet}. The effectiveness of RDB has been comprehensively examined \cite{liu2019griddehazenet,zhang2018residual}. In here, learning the residual information is beneficial to the color-mapping performance. The total of seven convolutional layers are used in RDB, where the first six layers aim at increasing the number of feature maps and the last layer concatenates all feature maps generated from these layers.
    At the end of RDB, a global context block is presented to encourage the network to learn the global color mapping, since local color mapping might introduce the degradation on the results due to the pixel misalignment between RAW and RGB image pairs. The reason is evident as the existence of misalignment misleads the neural network to map color into incorrect pixel locations.
     By considering that the convolutional kernel only covers the local information of an image, \cite{wang2018non} proposed a non-local attention mechanism. This work can realize the dependency between long-distance pixels so that the value at a query point can be calculated by the weighted sum of the features of all positions on the input feature. However, heavy computation is required, especially when the feature map has a large size (e.g., the full resolution input image from ZRR dataset). By experiments, \cite{cao2019gcnet} claims that the attention map obtained from different query points has minor differences. Therefore, they propose a lightweight global context block (GCB) that simplifies the non-local module and combines with the global context framework and the SE block \cite{hu2018squeeze}. The GCB encourages the network to learn key information spatial-wise and channel-wise while effectively reduce the computation complexity. These characteristics are exactly what we look for in this RAW-to-RGB mapping problem. 
    
	For up-sampling and down-sampling, we borrow the idea from the discrete wavelet transform (DWT), since the nature of DWT decomposes the input feature maps into the high-frequency and low-frequency components, in which the low-frequency one can be served as the result from average pooling (further discussion can be found in Section~\ref{dwtsection}). As shown in Fig.~\ref{fig:subblocks}, we use the low-frequency component as part of our down-sampling feature maps and connect the high-frequency part to the up-sampling block for image recovery (i.e., inverse DWT). However, the feature maps produced by frequency-domain operation might be lack of spatial correlation. Therefore, an additional spatial convolutional layer is adopted to downsample the feature map with learned kernels. Similarly, a pixel-shuffle operation along with a spatial convolutional layer is employed for up-sampling as the complement to the IDWT. The combination of frequency-domain and spatial-domain operations facilitates the learning of abundant features in up-sampling and down-sampling blocks. At the end of the proposed method, we use a Pyramid Pooling block \cite{chen2018encoder} to further enlarge the receptive field.
\subsection{Two-Branch Network}
By consolidating the encoder-decoder structure with previously mentioned modules, our network is able to surpass the state-of-the-art when trained on the RAW images. However, using multiple neural networks to train on different low-level vision tasks is a more effective way to learn image ISP. One of the reasons is that feeding distinct data to different network branches can provide abundant information during training. 
Recently, the two-stream design has been successfully applied in various computer vision tasks, especially in video field. Note that fusing the information from different formats of input (e.g., optical flow and image frames) can significantly improve the network performance. 
Inspired by \cite{carreira2017quo,feichtenhofer2016convolutional}, we build AWNet based on the idea of two-branch architecture to facilitate network performance on different low-level imaging tasks by utilizing different inputs. 
Our two-branch design contains two encoder-decoder models, namely the RAW model and the demosaiced model. In here, the RAW model is trained on $224\times224\times4$ RAW images, and the demosaiced branch takes $448\times448\times3$ demosaiced images as input. For the RAW model, there is a need to make the prediction size and ground truth size consistent. Therefore, this branch pays more attention to the recovery of high frequency details. For its counterpart, the demosaiced branch has no need to upscale the output size for consistency. Instead, this branch focuses more on the color mapping between the demosaiced image and RGB color image.
We train the two networks separately and average their predictions at testing. As expected, a great performance boost is observed by applying this architecture (see details in Section~\ref{section:ensemble}).

\subsection{Discrete Wavelet Transform}
\label{dwtsection}
To elaborate the reason of choosing DWT in our design opinion, we introduce the connection between DWT and traditional pooling operation. In 2D discrete wavelet transform, there are four filters, i.e., $f_{LL}$, $f_{LH}$, $f_{HL}$, and $f_{HH}$, can be used to decomposed an image \cite{mallat1989theory}. By convolving with each filter, a full-size image $x$ is split into 4 sub-bands, i.e., $x_{LL}$, $x_{LH}$, $x_{HL}$, and $x_{HH}$. Due to the nature of DWT, we can express $x_{LL}$ as $(f_{LL} \circledast x)\downarrow_{2}$ (the expressions of $x_{LH}$, $x_{HL}$, and $x_{HH}$ are similar), where $\circledast$ represents convolutional operation and $\downarrow_{2}$ indicates down-sampling by the scale factor of 2. According to the bi-orthogonal property, the original image $x$ can be restored by IDWT, i.e., $x = IDWT(x_{LL}, x_{LH}, x_{HL}, x_{HH})$. Therefore, the down-sampling and up-sampling operations of DWT can be considered as lossless. In addition, inspired by \cite{liu2018multi}, the wavelet transform can be employed to replace the traditional pooling operation that usually causes information loss. We define the mathematical format to further elaborate the connection between DWT and pooling operation. For example, in Haar DWT, $f_{LL}= \big(\begin{smallmatrix}
  1 & 1\\
  1 & 1
\end{smallmatrix}\big)$. Thus, the $(m,n)$-th value of $x_{LL}$ after 2D Haar wavelet transform can be defined as \\
\begin{equation}
\label{eqn:equation1}
    x_{LL}(m,n) = x(2m - 1, 2n - 1)+x(2m - 1, 2n)+x(2m, 2n - 1)+x(2m, 2n).
\end{equation}
Moreover, by defining $x_{p}$ to be the feature map after p-level of average pooling, the $(m,n)$-th value of $x_{p}$ can be expressed as 
\begin{equation}
\label{eqn:equation2}
\scriptstyle
\begin{split}
x_{p}(m,n) & = 0.25\times(x_{p-1}(2m - 1, 2n - 1) + x_{p-1}(2m - 1, 2n) \\ 
 & + x_{p-1}(2m, 2n - 1)+x_{p-1}(2m, 2n)).
\end{split}
\end{equation}
As we can see, Eq.~(\ref{eqn:equation2}) is highly correlated with Eq.~(\ref{eqn:equation1}). By taking four subbands into account, pooling operation discards all the high-frequency components and only makes use of low-frequency part. Therefore, the information loss in traditional pooling operation is severe. To alleviate this problem, we design our up-sampling and down-sampling modules in the way that uses both wavelet transform and convolutional operation to manage scaling. By doing that, our network can learn from both spatial and frequency information. Our experiments reveal the superior performance of this design (see details in Section~\ref{section:ablation}).

\subsection{Loss Function}
In this section, we introduce our three loss functions and the multi-scale loss strategy. We denote $I$ as the target RGB image and $\tilde{I}$ as the predicted result from our method.

\textbf{Pixel loss.}
We adopt the Charbonnier \cite{bruhn2005lucas,zhang2018image} loss as an approximate $L_{1}$ term for our loss function to better handle outliers and improve the performance. From previous experiments, we realize that Charbonnier loss can efficiently improve the performance on the signal-to-noise ratio of reconstructed images. In addition, Charbonnier loss has been applied in multiple image reconstruction tasks and outperforms the traditional $L_{2}$ penalty \cite{zhang2018image}. The Charbonnier penalty function is defined as:
	\begin{equation}
	L_{char} = \sqrt{(\tilde{I} - I)^2 + \epsilon^2},
	\end{equation}
where we set $\epsilon$ to $1e-3$. Note that using only the pixel loss on RAW-to-RGB mapping results in blurry images as reported in \cite{uhm2019w}. Thus, we redeem this problem by adding other feature loss functions.

\textbf{Perceptual loss.} To deal with the pixel misalignment problem from ZRR dataset, we also employ perceptual loss. The loss function is defined as
	\begin{equation}
	L_{P} = L_{MSE}(F(\tilde{I}) - F(I)),
	\end{equation}
where $F$ denotes the pretrained VGG-19 network, $\tilde{I}$ and $I$ represent the predicted image and ground truth, respectively. 
As misaligned images are processed by the pretrained VGG network, the resulting downsampled feature maps have fewer variants in terms of the misalignment. Therefore, adding a $L_{2}$ term on such feature maps is beneficial for the network to recognize the global information and minimize the perceptual difference between the reconstructed image and the ground truth image.

\textbf{SSIM loss.}
We also employ the structural similarity (SSIM) loss $L_{SSIM}$ \cite{wang2003multiscale} that is aiming to reconstruct the RGB images by enhancing on structural similarity index. The resulting images are more perceptually accepted than the predictions without applying SSIM loss. Note that the loss function can be defined as:
\begin{equation}
	L_{SSIM} = 1 - F_{SSIM}(\tilde{I} - I),
\end{equation}
where $F$ denotes the function of calculating structural similarity index. \\ 

\textbf{Multi-scale loss function.}
Inspired by \cite{qian2018attentive}, we apply supervision on outputs from different decoder layers to refine reconstructed images of different sizes. For each scale level, we focus on different restoration aspects, thus different loss combinations are applied. In our RAW model, there are 5 up-sampling operations, which form feature maps in 6 different scales, named as scale 1-6 from small to large. Similarly, there are 5 different scales presented in the demosaiced model and we name those as scales 1-5.

1). Scale 1-2 process feature maps that are down-scaled by a factor of 16 and 32. The feature maps at this scale contain less context information compare with ground truth. Thus, we mainly focus on global color and tone mapping. These layers are supervised only by Charbonnier loss, which can be written as:
\begin{equation}
	L_{1,2} = L_{char}.
\end{equation}

2). Scale 3-4 are computed on feature maps with down-scaled factors of 4 and 8; since these features are smaller as compared to the size of ground truth yet contain richer information than the scale 1-2, we apply a loss combination that incorporates perceptual and Charbonnier losses to perform global mapping while remaining the perceptual acceptance. The loss function of these layers is defined as:
\begin{equation}
	L_{3,4} = L_{char} + 0.25 \times L_{P}.
\end{equation}

3) In scale 5-6, the size of feature maps is close or equal to the original one, thus we are able to pay more attention to the recovery of image context in addition to the color mapping. We choose a more comprehensive loss combination at this level, which can be shown as:
\begin{equation}
	L_{5,6} = L_{char} + 0.25 \times L_{P} + 0.05 \times L_{SSIM}.
\end{equation}
Note that we manually choose the coefficients of different loss terms. The total loss function can be expressed as:
\begin{equation}
    L_{total} = \sum_{n=1}^{k} L_{n},
\end{equation}
where $k$ is equal to 5 and 6 for demosaiced model and RAW model, respectively.
\begin{figure*}[!htb]
\centering
\begin{minipage}[h]{0.2\linewidth}
    \centering
    \includegraphics[width=\linewidth]{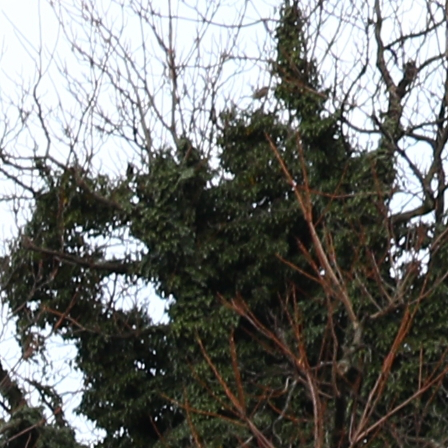}
    \centerline{\scriptsize PSNR / SSIM}
    \includegraphics[width=\linewidth]{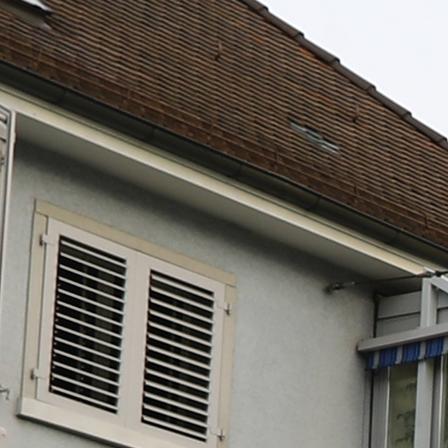}
    \centerline{\scriptsize PSNR / SSIM}
    \includegraphics[width=\linewidth]{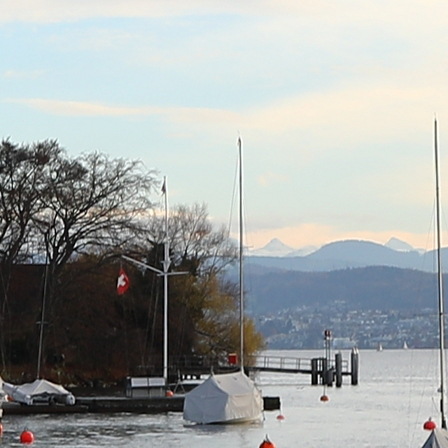}
    \centerline{\scriptsize PSNR / SSIM}
    \includegraphics[width=\linewidth]{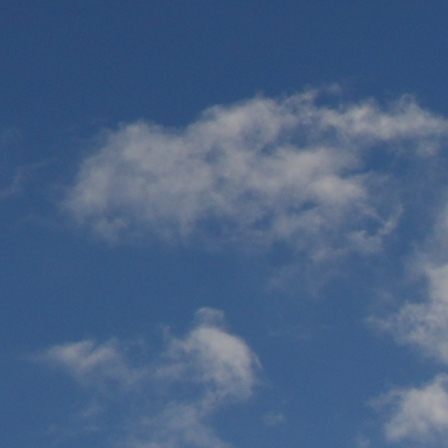}
    \centerline{\scriptsize PSNR / SSIM}
    \centerline{\scriptsize Ground Truth}
    \vspace{1mm}
\end{minipage}
\hspace{3mm}
\begin{minipage}[h]{0.2\linewidth}
    \centering
    \includegraphics[width=\linewidth]{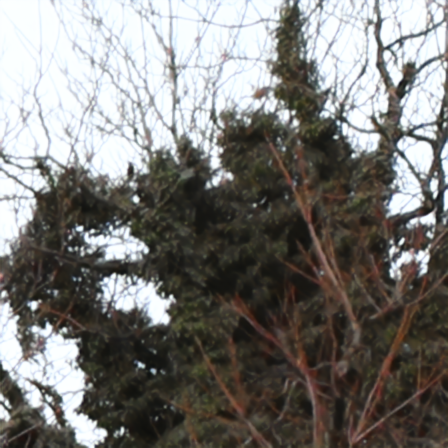}
    \centerline{\scriptsize 21.9890 / 0.7534}
    \includegraphics[width=\linewidth]{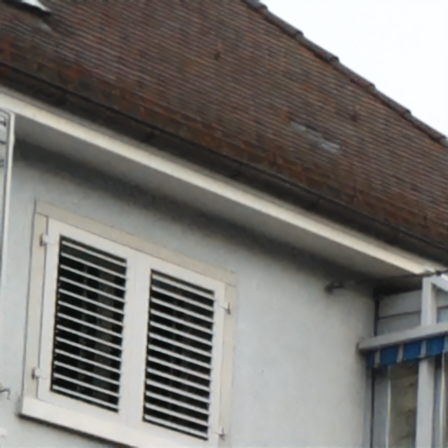}
    \centerline{\scriptsize 20.0099 / 0.6912}
    \includegraphics[width=\linewidth]{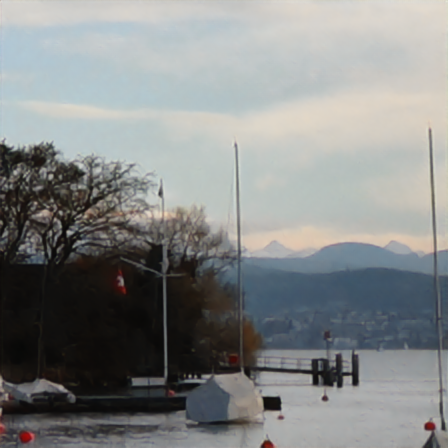}
    \centerline{\scriptsize 16.8522 / 0.7124}
    \includegraphics[width=\linewidth]{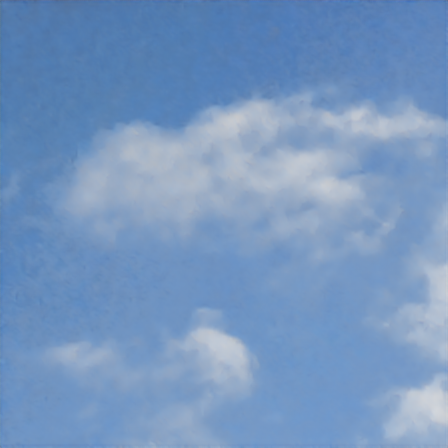}
    \centerline{\scriptsize 13.6305 / 0.8503}
    \centerline{\scriptsize Ours-3 Output}
    \vspace{1mm}
\end{minipage}
\hspace{3mm}
\begin{minipage}[h]{0.2\linewidth}
    \centering
    \includegraphics[width=\linewidth]{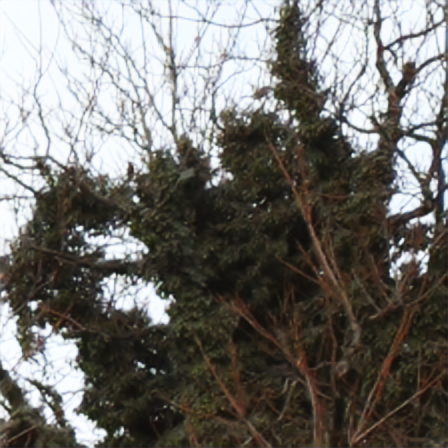}
    \centerline{\scriptsize 23.0414 / 0.7743}
    \includegraphics[width=\linewidth]{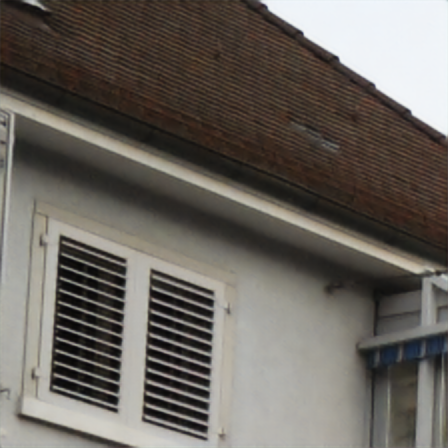}
    \centerline{\scriptsize 21.8351 / 0.7082}
    \includegraphics[width=\linewidth]{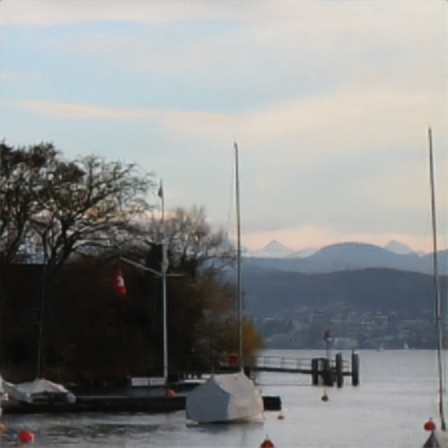}
    \centerline{\scriptsize 17.7539 / 0.7195}
    \includegraphics[width=\linewidth]{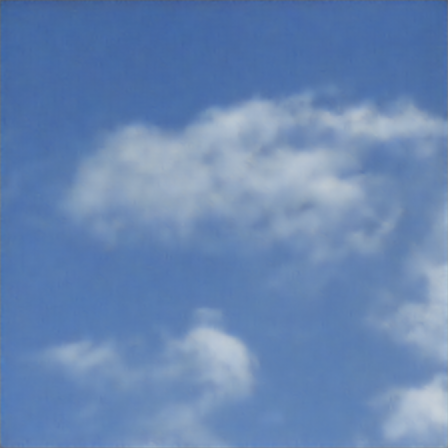}
    \centerline{\scriptsize 16.4578 / 0.8939}
    \centerline{\scriptsize Ours-4 Output}
    \vspace{1mm}
\end{minipage}
\hspace{3mm}
\begin{minipage}[h]{0.2\linewidth}
    \centering
    \includegraphics[width=\linewidth]{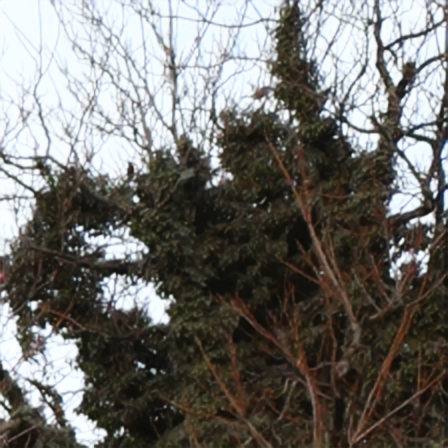}
    \centerline{\scriptsize 23.2214 / 0.7981}
    \includegraphics[width=\linewidth]{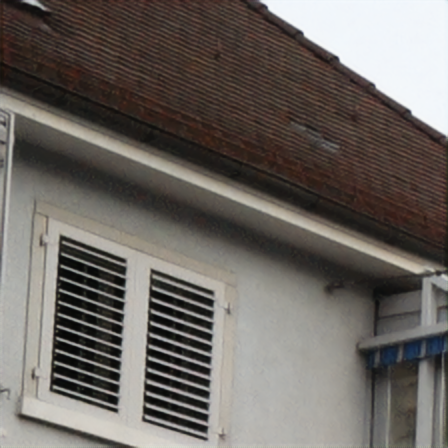}
    \centerline{\scriptsize 21.9083 / 0.7138}
    \includegraphics[width=\linewidth]{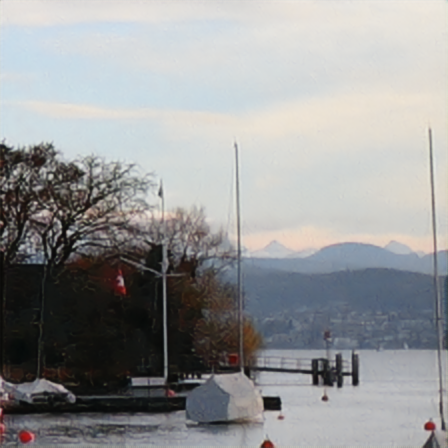}
    \centerline{\scriptsize 20.1982 / 0.7381} 
    \includegraphics[width=\linewidth]{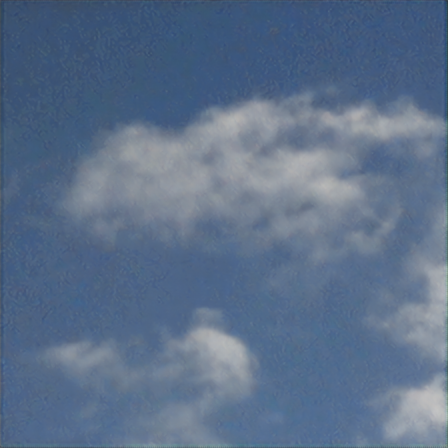}
    \centerline{\scriptsize 24.4593 / 0.9016}
    \centerline{\scriptsize Ensemble Output}
    \vspace{1mm}
\end{minipage}
\caption{PSNR/SSIM and visual comparisons of reconstructed images from different network models. Ours-3 and Ours-4 denote our demosaiced and RAW models, respectively. Zoom-in for better views.}
\label{fig:output}
\vspace{0.2in}
\end{figure*}

\section{Experiments}
We conduct comprehensive experiments to demonstrate that the proposed method performs favorable against the baseline model \cite{ignatov2020replacing} in terms of quantitative and qualitative comparisons on ZRR dataset.

\subsection{Datasets}
To enhance smartphone images, the Zurich dataset from AIM 2020 Learned Smartphone ISP Challenge \cite{ignatov2020replacing} provides 48043 RAW-RGB image pairs (of size $448\times448\times1$ and $448\times448\times3$, respectively). The training data has been divided into 46,839 image pairs for training and 1,204 ones for testing. In addition, 168 full resolution image pairs are used for perceptual validation. For data preprocessing and augmentation, we normalize the input data and perform vertical and horizontal flipping.\\ 

\subsection{Training Details}
Our model is trained on PyTorch framework with Intel i7, 32GB of RAM, and two NVIDIA RTX2080 Ti GPUs. The batch size is set to 6 and 2 for the RAW model and the demosaiced model, respectively. Except for that, our two models share the same training strategy. We employ Adam optimizer \cite{kingma2014adam} with $\beta_{1} = 0.9, \beta_{2} = 0.999$ and set the initial learning rate as $1\times10^{-4}$. We decrease the learning rate by half in every 10 epochs and train for 50 epochs in total.\\

\begin{table}
\centering
\vspace{-4mm}
\caption{Validation scores by different model ensembles. We use red text to indicate the best
performance and blue text to indicate the second best performance.}
\vspace{3mm}
\begin{tabular}{|c||c||c|}
\hline
\begin{tabular}[c]{@{}c@{}}RAW model\\ PSNR (dB) / SSIM \end{tabular} & \begin{tabular}[c]{@{}c@{}}Demosaiced model\\ PSNR (dB) / SSIM \end{tabular} & \begin{tabular}[c]{@{}c@{}}Ensemble Score\\ PSNR (dB) / SSIM \end{tabular}  \\ 
\hline
21.36 / 0.7429                                                        & 21.30 / 0.7455                                                               & 21.60 / \textcolor{red}{0.7818}                                             \\
21.36 / 0.7429                                                        & 21.38 / 0.7522                                                               & 21.92 / 0.7761                                                              \\
21.36 / 0.7429                                                        & 21.52 / 0.7484                                                               & \textcolor{blue}{21.95} / \textcolor{blue}{0.7788}                          \\
21.36 / 0.7429                                                        & 21.58 / 0.7488                                                               & 21.79 / \textcolor{red}{0.7818}                                             \\
21.38 / 0.7451                                                        & 21.58 / 0.7488                                                               & \textcolor{red}{21.97} / 0.7784                                             \\
\hline
\end{tabular}

\label{tab:table1}
\end{table}

\begin{table}
\centering

\caption{\label{tab:challenge}The result of AIM2020 Learned Smartphone ISP Challenge for the two tracks. Our method can achieve high MOS while remaining competetive in PSNR and SSIM metrics.}
\vspace{3mm}
\begin{tabular}{|c||c|c|c|c|c|c|c|} 
\hline
     & \multicolumn{3}{|c|}{Track 1}                                                                       & \multicolumn{4}{|c|}{Track 2}                                                                       \\ 
\hline
Rank & Method & PSNR                                         & SSIM                                       & Method & PSNR                                        & SSIM & MOS                                        \\ 
\hline
$1$    & Airia\_CG      & \textcolor[rgb]{0.133,0.133,0.133}{22.2574}  & \textcolor[rgb]{0.133,0.133,0.133}{0.7913} &  MW-ISPNet      & \textcolor[rgb]{0.133,0.133,0.133}{21.574}  & \textcolor[rgb]{0.133,0.133,0.133}{0.777} & 4.7 \\ 
\hline
$2$    & skyb      & \textcolor[rgb]{0.133,0.133,0.133}{21.9263} & \textcolor[rgb]{0.133,0.133,0.133}{0.7865} & \textbf{AWNet}  & \textbf{\textcolor[rgb]{0.133,0.133,0.133}{21.861}}  & \textbf{0.7807} & \textbf{4.5}  \\ 
\hline
$3$    & MW-ISPNet      & \textcolor[rgb]{0.133,0.133,0.133}{21.9149}  & \textcolor[rgb]{0.133,0.133,0.133}{0.7842} & Baidu      & \textcolor[rgb]{0.133,0.133,0.133}{21.9089} & 0.7829 & 4.0 \\ 
\hline
$4$    & Baidu       & \textcolor[rgb]{0.133,0.133,0.133}{21.9089}  & \textcolor[rgb]{0.133,0.133,0.133}{0.7829} & skyb      & \textcolor[rgb]{0.133,0.133,0.133}{21.734}  & \textcolor[rgb]{0.133,0.133,0.133}{0.7891} & 3.8 \\ 
\hline
$5$   & \textbf{AWNet}  & \textbf{\textcolor[rgb]{0.133,0.133,0.133}{21.8610}}    & \textbf{0.7807}                                    & STAIR      & \textcolor[rgb]{0.133,0.133,0.133}{21.569}  & \textcolor[rgb]{0.133,0.133,0.133}{0.7846} & 3.5 \\
\hline

\end{tabular}
\end{table}

\subsection{Ensemble Strategy}
\label{section:ensemble}
Inspired by \cite{timofte2016seven}, we applied a self-ensemble mechanism during the validation and testing stage of AIM2020 Learned Smartphone ISP Challenge. Specifically, we use ensembles comprised of 8 variants (original, rotated $90^{\circ}$, rotated $180^{\circ}$, rotated $270^{\circ}$, rotated $90^{\circ}$ \& flipped, rotated $180^{\circ}$ \& flipped, and rotated $270^{\circ}$ \& flipped ones). After that, we average out the ensemble outputs and obtain our final result.
To evaluate the benefit of ensembles, we apply our method to the validation dataset (without ground truth) during the development stage to validate our methods by calculating the PSNR values. In our experiments, the non-ensembles version of the RAW model and the demosaiced model in Track 1 achieves 21.55 dB and 21.68 dB on the validation dataset (without ground truth), respectively. Subsequently, by averaging out the results from both models, the PSNR can be significantly boosted to 21.97 dB. To achieve optimal ensemble result, for each model, we prepare weights with different PSNR scores, and then carry out experiments to test different combinations of weights across two models (see Table~\ref{tab:table1} for details).
 At the final testing stage, we choose the 21.36 dB (RAW model) and 21.52 dB (demosaiced model) weights to generate predictions. Fig.~\ref{fig:output} shows the qualitative and quantitative results from these models and their ensemble outcomes (tested on offline validation data from provided ZRR dataset). Table~\ref{tab:challenge} shows the result of AIM2020 Learned Smartphone ISP Challenge \cite{ignatov2020aim_ISP} for the two tracks. We are ranked in the $5^{th}$ and $2^{nd}$ place in track 1 and 2, respectively.

\begin{figure*}[!htb]
\begin{minipage}[htb]{0.16\linewidth}
    \centering
    {\includegraphics[width=\linewidth]{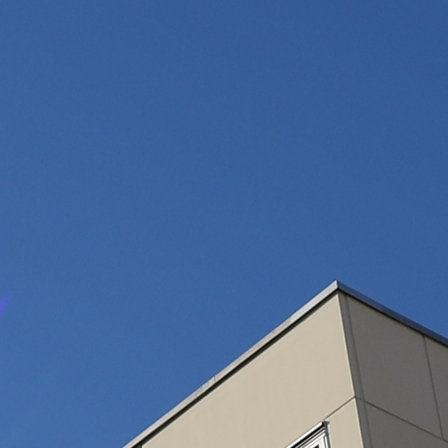}}
    {\includegraphics[width=\linewidth]{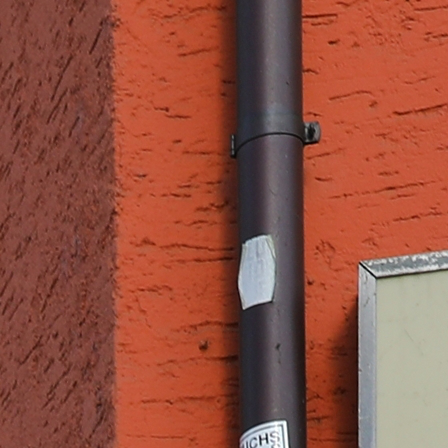}}
    {\includegraphics[width=\linewidth]{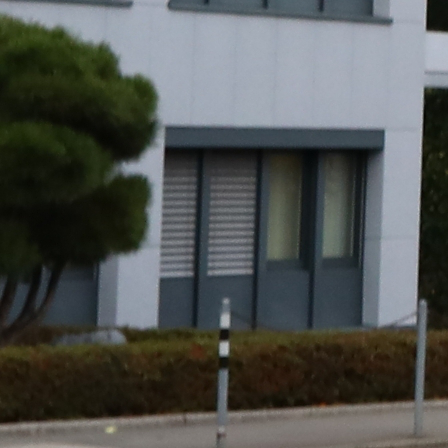}}
    {\includegraphics[width=\linewidth]{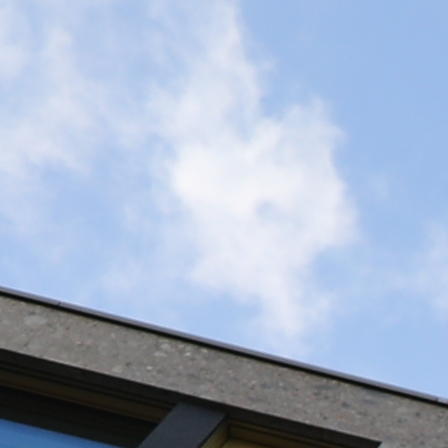}}
    {\includegraphics[width=\linewidth]{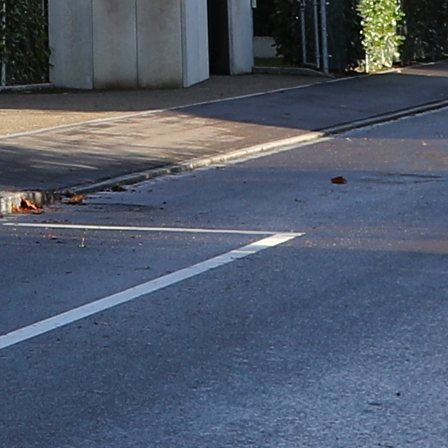}}
    \centerline{Ground Truth}
    \vspace{0.0001cm}
\end{minipage}
\begin{minipage}[htb]{0.16\linewidth}
    \centering
    {\includegraphics[width=\linewidth]{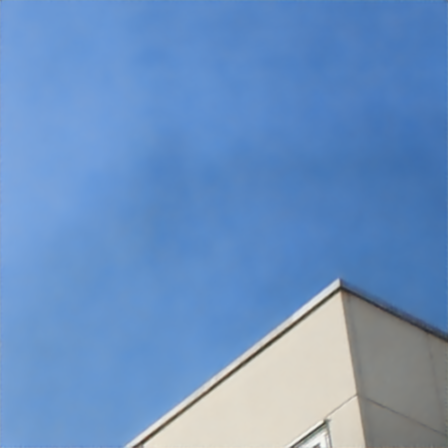}}
    {\includegraphics[width=\linewidth]{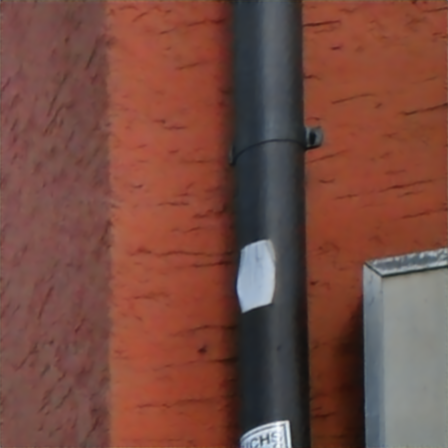}}
    {\includegraphics[width=\linewidth]{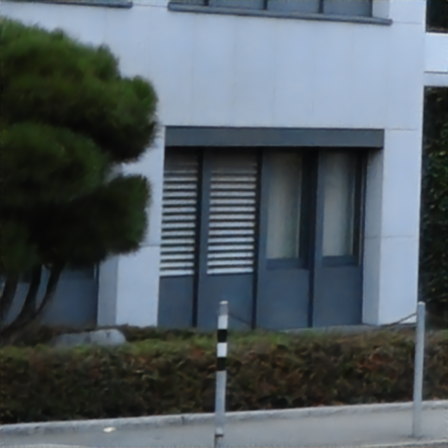}}
    {\includegraphics[width=\linewidth]{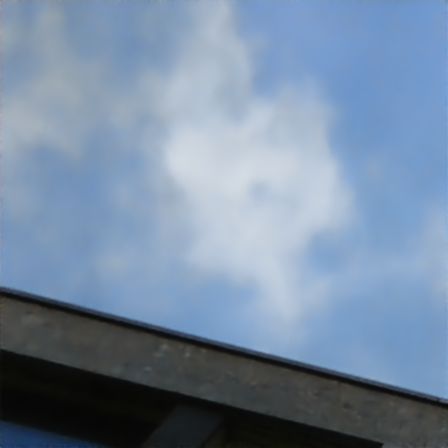}}
    {\includegraphics[width=\linewidth]{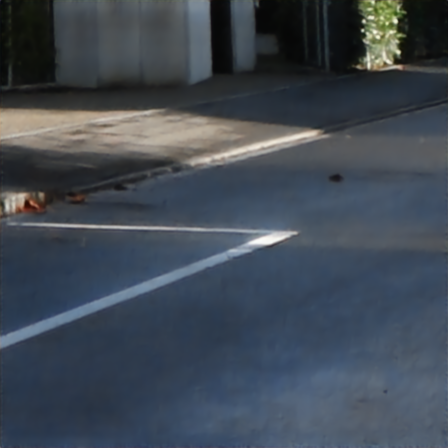}}
    \centerline{U-Net}
    \vspace{0.0001cm}
\end{minipage}
\begin{minipage}[htb]{0.16\linewidth}
    \centering
    {\includegraphics[width=\linewidth]{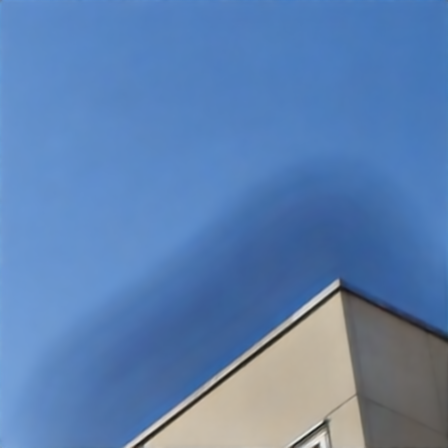}}
    {\includegraphics[width=\linewidth]{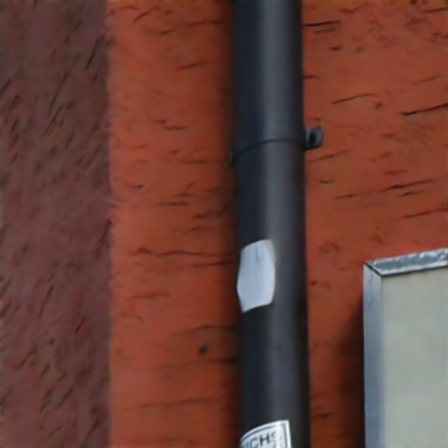}}
    {\includegraphics[width=\linewidth]{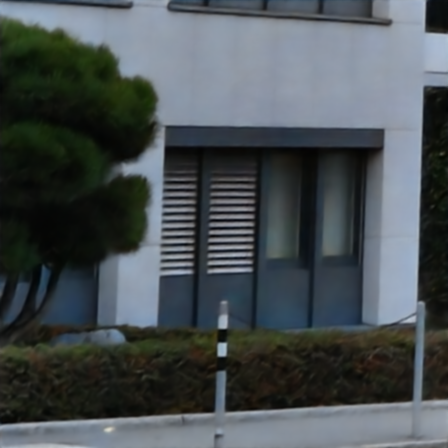}}
    {\includegraphics[width=\linewidth]{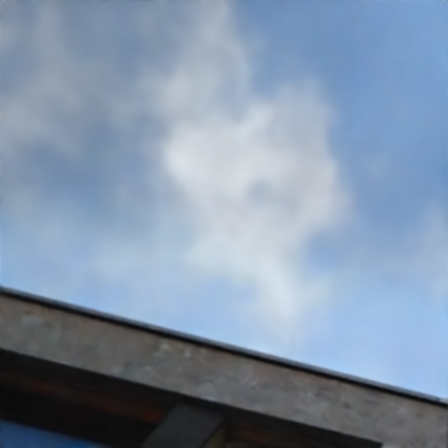}}
    {\includegraphics[width=\linewidth]{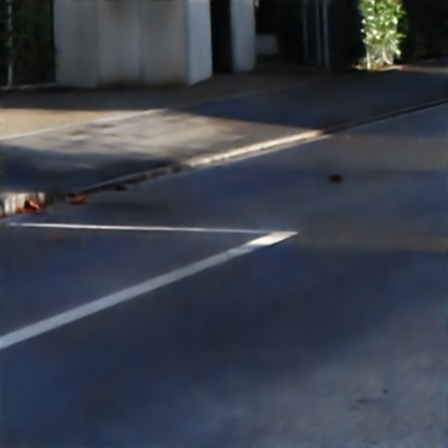}}
    \centerline{RCAN}
    \vspace{0.0001cm}
\end{minipage}
\begin{minipage}[htb]{0.16\linewidth}
    \centering
    {\includegraphics[width=\linewidth]{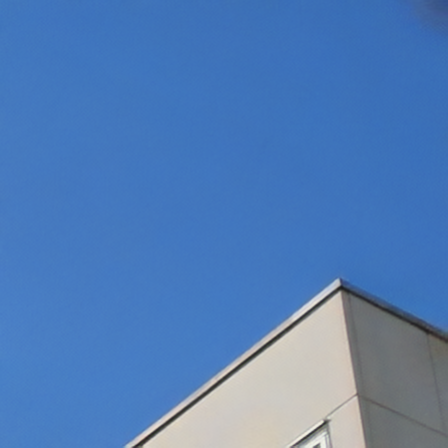}}
    {\includegraphics[width=\linewidth]{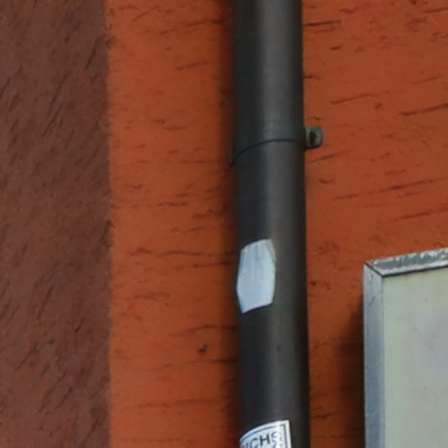}}
    {\includegraphics[width=\linewidth]{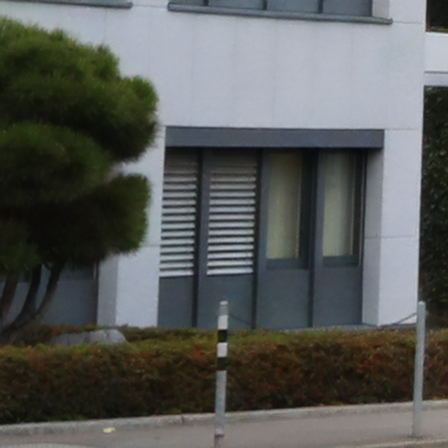}}
    {\includegraphics[width=\linewidth]{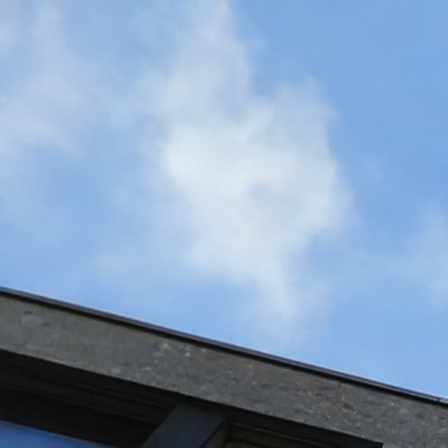}}
    {\includegraphics[width=\linewidth]{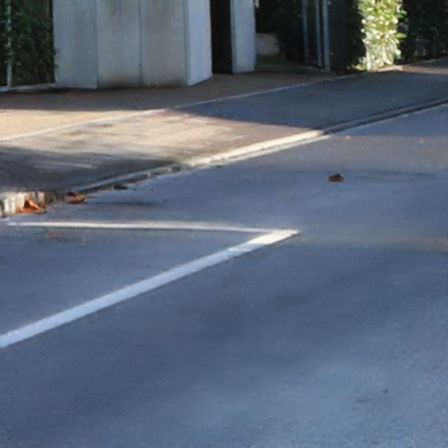}}
    \centerline{PyNet}
    \vspace{0.0001cm}
\end{minipage}
\begin{minipage}[htb]{0.16\linewidth}
    \centering
    {\includegraphics[width=\linewidth]{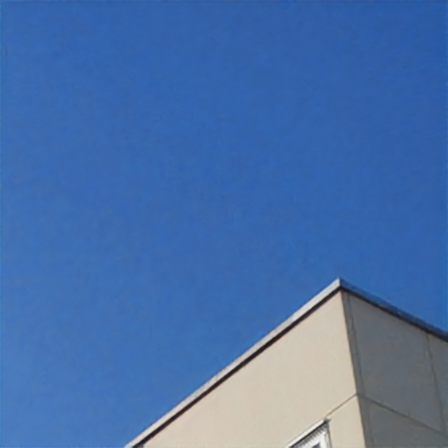}}
    {\includegraphics[width=\linewidth]{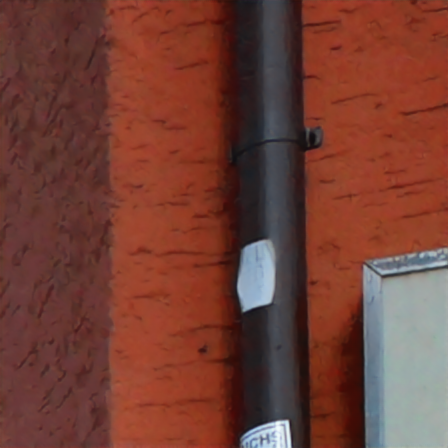}}
    {\includegraphics[width=\linewidth]{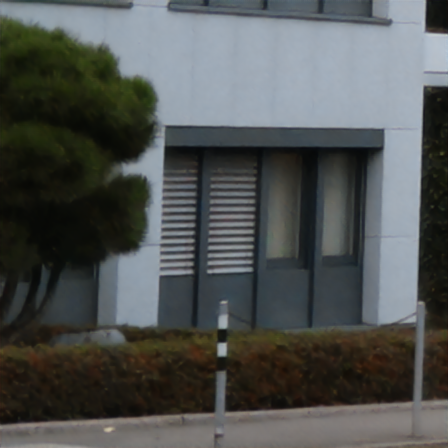}}
    {\includegraphics[width=\linewidth]{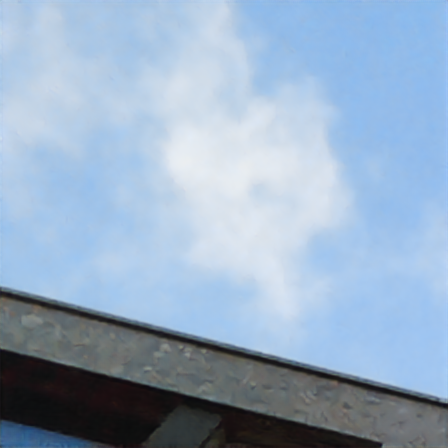}}
    {\includegraphics[width=\linewidth]{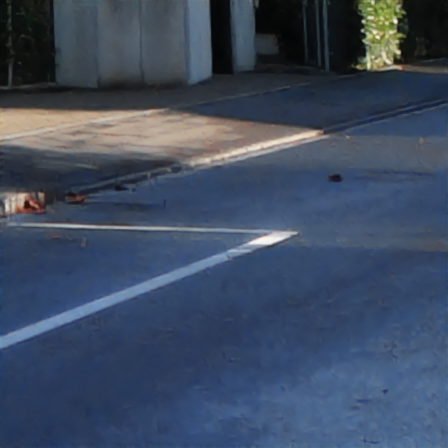}}
    \centerline{Ours-3}
    \vspace{0.0001cm}
\end{minipage}
\begin{minipage}[htb]{0.16\linewidth}
    \centering
    {\includegraphics[width=\linewidth]{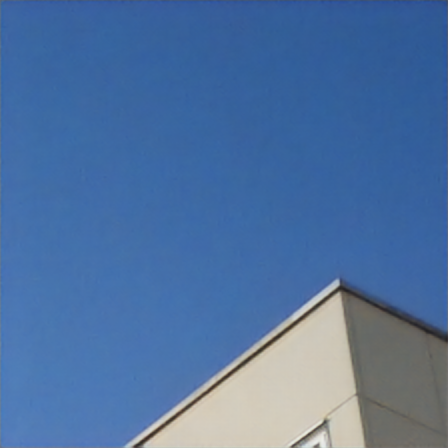}}
    {\includegraphics[width=\linewidth]{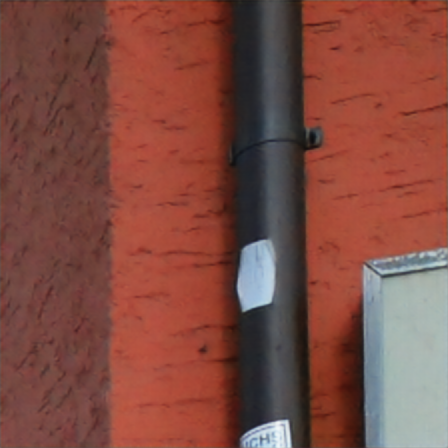}}
    {\includegraphics[width=\linewidth]{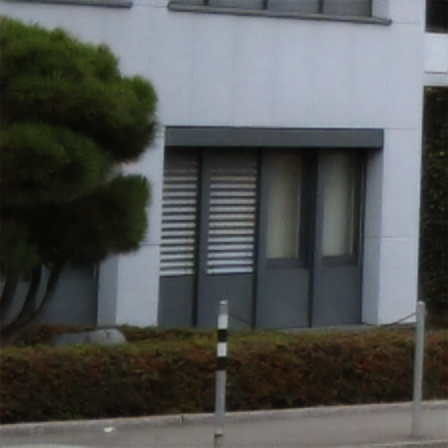}}
    {\includegraphics[width=\linewidth]{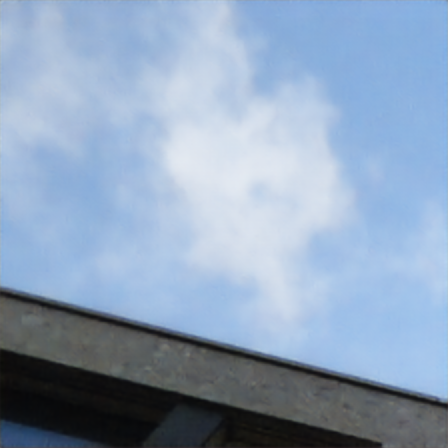}}
    {\includegraphics[width=\linewidth]{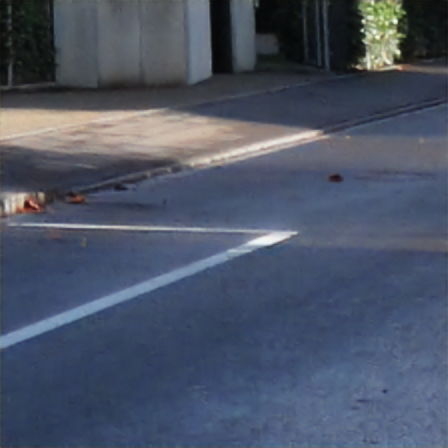}}
    \centerline{Ours-4}
    \vspace{0.0001cm}
\end{minipage}
\vspace{1mm}
\caption{Qualitative comparisons of reconstructed images from different networks. Ours-3 and Ours-4 denote our demosaiced and RAW models, respectively. Zoom-in for better views.}
\label{fig:qualitative}
\vspace{-2mm}
\end{figure*}

\subsection{Performance Comparisons and Ablation Studies}
\label{section:ablation}
We conduct an experiment by first comparing it with other state-of-the-arts to demonstrate the superior performance of our method. After that, we provide solid justification for the effectiveness of wavelet transform and global context blocks. Our proposed method is tested on offline validation data that is provided during the development stage. We choose some popular network architectures from different computer vision tasks, including UNet and RCAN, for comparisons. The qualitative comparisons can be seen from Table~\ref{tab:table2}, and Fig.~\ref{fig:qualitative} shows the qualitative comparison between our method and other state-of-the-arts. As we can see, both U-Net and RCAN have some color mapping artifacts, which manifests the incapability of mapping color into RGB space correctly in a pixel-to-pixel manner. For example, in the first row of Fig.~\ref{fig:qualitative}, the color of the sky is inaccurately predicted. Although the PyNet performs better in the color mapping aspect, it tends to obscure the image details. This artifact is obvious in the second, the third, and the last row of images. Beneficial from DWT and GCB blocks, the proposed method remedies these artifacts, which present in other state-of-the-arts. Moreover, the RAW model provides more fine image details whereas the demosaiced model has a better matching in color space; this reveals the effectiveness of our design.
\begin{table}[!htb]
\vspace{-2mm}
	\centering
	    \caption{\label{tab:table2}Quantitative results from different models. Both of our proposed models outperform the state-of-the-arts. Ours-3 and Ours-4 indicate our demosaiced and RAW models, respectively.}
	    \vspace{3mm}
        \begin{tabular}{|c||c|} 
        \hline
        Models & PSNR (dB) / SSIM  \\ 
        \hline
        U-Net  & 21.01 / \textbf{0.7520}    \\ 
        \hline
        RCAN   & 20.85 / 0.7510    \\ 
        \hline
        PyNet  & 21.17 / 0.7460    \\ 
        \hline
        Ours-3 & \textbf{21.58} / 0.7488    \\ 
        \hline
        Ours-4 & 21.38 / 0.7451    \\
        \hline

        \end{tabular}

\vspace{-2mm}
\end{table}

To validate that the wavelet transform and GCB blocks enable to improve the output performance, two corresponding experiments are conducted. The first one is to remove wavelet transform and GCB blocks (see Fig.~\ref{fig:subblocks}) from residual wavelet up-sampling module, residual wavelet down-sampling module, and global context res-dense module; the another one is to restore GCB blocks and leave wavelet transform blocks absent. As shown in Table~\ref{tab:ablation}, by adding GCB blocks, both of our models can be boosted by 0.1 dB in terms of PSNR metric. The performance can be further improved by 0.2 dB while adding DWT block. Note that all these variants are trained in the same way as before and tested on the offline validation dataset from AIM2020 Learned Smartphone ISP Challenge.
\begin{table}
\vspace{-2mm}
\centering
\caption{The benefit of using DWT and GCB blocks is evident. Both of our models can receive approximate 0.3 dB gains.}
\vspace{3mm}
\begin{tabular}{|c||c||c|} 
\hline
Model                                                                            & Operation                                         & PSNR (dB) \textbackslash{} SSIM  \\ 
\hline
\multirow{3}{*}{\begin{tabular}[c]{@{}c@{}}Demosaiced model\end{tabular}} & w/o DWT and w/o GCB & 21.13 / 0.7398                   \\ 
\cline{2-3}
 & w/o DWT & 21.22 / 0.7421\\ 
\cline{2-3}
 & proposed model   & \textbf{21.38 / 0.7451}\\ 
\hline
\multirow{3}{*}{\begin{tabular}[c]{@{}c@{}}RAW model\end{tabular}} & w/o DWT and w/o GCB & 21.22 / 0.7325~                  \\ 
\cline{2-3}
& w/o DWT & 21.31 / 0.7398\\ 
\cline{2-3}
& proposed model    & \textbf{21.58 / 0.7488}\\
\hline
\end{tabular}

\label{tab:ablation}
\vspace{-2mm}
\end{table}

Our qualitative and quantitative results validate superiority of our two-branch design as well as the effectiveness of wavelet transform block and attention mechanism, in the application of learning RAW-to-RGB color mapping.

\section{Conclusion}
In this paper, we propose a novel two-branch network structure, named AWNet, which can effectively enhance the smartphone images.
We embed wavelet transform blocks into the scaling modules associated with convolutional operations that enable our network to learn from both the spatial and frequency domains.
In addition, the presence of GCB blocks improves the robustness of our network to deal with the misalignments that occurred in the ZRR dataset.
Our work can shed some light on the application of wavelet transform in image ISP problem.
As for future work, our network is able to tackle other low-level imaging tasks, such as image denoising and super-resolution.

\bibliographystyle{splncs04}
\bibliography{egbib}
\end{document}